\documentstyle[amssymb,preprint,aps]{revtex}
\tightenlines

\begin{document}
{\Large\bf QUASI FREE ELECTROFISSION of $^{238}$U.}

\begin{center}
\vspace{0.3cm}
V. P. Likhachev$^{a}$ J. Mesa$^{a,b}$ J. D. T. Arruda-Neto$^{a,c}$, B. V.\
Carlson$^{d}$, A. Deppman$^{a}$, \\M. S. Hussein$^{a}$, V. O. Nesterenko$^{e}$%
, F. Garcia$^{f}$ and O. Rodriguez$^{b}$.
\end{center}

\vspace{0.3cm}
\noindent
{\it
$^{a}$ Instituto de F\'{i}sica, Universidade de S\~{a}o Paulo, S\~{a}o
Paulo, Brazil.\\
$^{b}$ Instituto Superior de Ciencias y Tecnologia Nucleares, Havana, Cuba.\\
$^{c}$ Universidade de Santo Amaro, S\~{a}o Paulo, Brazil.\\
$^{d}$ Instituto de Estudos Avan\c{c}ados-Centro T\'{e}cnico Aeroespacial,\\
S\~{a}o Jos\'{e} dos Campos, Brazil.\\
$^{e}$ Bogolubov Laboratory of Theoretical Physics, JINR, Dubna, Russia.\\
$^{f}$ Universidade Estadual de Santa Cruz, Ilheus, Bahia, Brazil.
}
\begin{abstract}
We present the result of a theoretical study of the quasi free
electrofission of $^{238}$U. The exclusive differential cross sections for
the quasi free scattering reaction stage have been calculated in PWIA, using
a Macroscopic-Microscopic approach for the description of the proton bound
states. The nuclear shape was parametrized in terms of cassinian ovaloids.
The equilibrium deformation parameters have been calculated by minimizing
the total nuclear energy. In the calculation the axially deformed
Woods-Saxon single particle potential was used. The obtained single particle
momentum distributions were averaged over the nuclear symmetry axis
direction. The occupation numbers were calculated in the BCS approach. The
fissility for the single hole excited states of the residual nucleus $^{237}$%
Pa was calculated on the statistical theory grounds\ both without taking
into account the pre-equilibrium emission of the particle, and with
preequilibrium emission in the framework of the exciton model.
\end{abstract}

\section{INTRODUCTION}

\bigskip

Quasi free scattering of high energy electrons (QF) on nuclei is the field
of nuclear physics which is traditionally devoted to the study of the single
particle aspects of nuclear structure: single particle binding energies,
momentum distributions, occupation numbers, etc\cite{Cardman}.

A new branch of these investigations \ is the study of decay channels of
single hole states in the residual nucleus, created as a result of the QF
process. Especially interesting is to study a fission decay following a QF
process. In this case we have a single particle process in the first
reaction stage, and essentially a collective process in the final reaction
stage. The collective degrees of freedom are excited in the intermediate
reaction stage due the residual interaction.

This is a new sort of nuclear reaction which may allow one to get unique
information on the dissociation of well defined single hole configurations (
which we can select by coincidence\ $\ (e,e^{\prime }p)$) into complex
nuclear configurations, and its role in nuclear fission. The new and most
important aspect of this reaction is that, after knocking-out a proton, we
obtain the heavy nucleus $^{237}$Pa in a single hole doorway state (see
discussion below) which could undergo \ nuclear fission. Indeed, instead of
dealing with collective doorway states, which are coherent sums of a great
number of 1p-1h configurations ( as the well-known giant resonances), these
non-collective doorway states will be represented by only one, well defined,
1h configuration. The residual interaction in $^{237}$Pa mixes these 1h
configurations into more complicate 2h-1p and 3h-2p ones. So, there would be
some competing channels for fission. It may occur either directly from 1h
configurations, or, with some delay, from mixed states (or their
components). In a QF process we have in the initial state only one
configuration; thus, the fission probability $P_{f}$ should be more
sensitive to the individual structure of this initial state as compared with
conventional reactions, where the effects of the structure are averaged out
over many single particle states forming the doorway.

\bigskip 
The unambiguous extraction of single hole contributions is possible
only in an exclusive experimental scheme (reaction $(e,e^{\prime }$ $pf)$)
and involves extremely thin targets (fission fragments have to leave the
target with small energy losses), high energy resolution, and coincidence
requirement between the final particles in order to separate the single hole
states. The exclusive $(e,e^{\prime }pf)$-experiment is very difficult for
practical realization, and never has been so far performed. The integral
contribution of the quasi free electron scattering to the fission process
was studied only in inclusive experiments: $(e,f)$ \cite{Likha} and $%
(e,e^{\prime }f)$) \cite{Hansen}. These works dealt only with the issue of
the QF contribution in electrofission.

The advent of high energy, CW, electron accelerators combined with the
development of high resolution facilities, opens the\ possibility of
studying the fission channel for quasi free electron scattering in an
exclusive experimental setup. The most suitable accelerator for this
experiment is at the Thomas Jefferson National Accelerator Facility (TJNAF).

For excitation of the residual nucleus to an well defined single hole state,
the initial and final state interactions have to be negligible. This
situation corresponds to high momentum transfer and high proton exit
energy,\ when the Plane Wave Impulse Approximation (PWIA) for the
calculation of the quasi free electron scattering cross section is valid.

This work presents the results of a PWIA \ calculations for the quasi free $%
(e,e^{\prime }p)$-differential cross section for deformed orbitals of $\
^{238}$U, in the framework of the macroscopic-microscopic approach,\ plus an
estimate of the fissility \ for single hole states in the residual nucleus $%
^{237}$Pa, performed on the statistical theory grounds. These \ calculations
could serve as first order magnitude guide-line for expected cross sections.

\bigskip

\section{PWIA CROSS SECTION}

\bigskip

\bigskip 
In the first order Born approximation the electron with initial
four-momentum $k_{1\mu }=(\overrightarrow{k}_{1},i\varepsilon _{1})$ and
final \ $k_{2\mu }=(\overrightarrow{k}_{2},i\varepsilon _{2})$ , \
transfers\ \ a virtual photon with four-momentum $q_{\mu }=(\overrightarrow{q%
},i\omega )=k_{1\mu }-k_{2\mu }$ ,\ resulting \ in the final state \ a
knocked-out \ nucleon \ with $p_{\mu }=(\overrightarrow{p_{p}},iE_{p})$ \
and \ a residual nucleus with $P_{A-1\mu }=(\overrightarrow{P}%
_{A-1},iE_{A-1})$.

In the impulse approximation \ a virtual photon interacts with a bound
nucleon (proton or neutron) of four-momentum $p_{m}=$\ $(\overrightarrow{p}%
_{m},iE_{m}),$which exits the nucleus with \ four-momentum $p_{\mu }=(%
\overrightarrow{p_{p}},iE_{p})$ \ without further interaction (no FSI). The
corresponding momentum diagram in the impulse approximation \ is shown in
Fig. 1 for the Laboratory system..

In the plane wave impulse approximation (PWIA) \ $\overrightarrow{p}_{m}=-%
\overrightarrow{P}_{A-1},$ \ and the missing quantities (momentum and energy
of the proton before interaction) can be defined from the energy and
momentum conservation law in the following way:

\begin{eqnarray}
\overrightarrow{p}_{m} &=&\overrightarrow{p}_{p}-\overrightarrow{q},
\label{1} \\
E_{m} &=&\omega -T_{p}-T_{A-1},  \nonumber
\end{eqnarray}
where $\ E_{m}=M_{A-1}+m_{p}-M_{A}$\ is the proton missing (or separation)
energy, $T_{p}$ is the kinetic energy of the outgoing proton, and $T_{A-1}$
is the kinetic energy of the residual nucleus. The momentum and energy
transfer of the virtual photon can be varied independently.

In the PWIA the six folded differential cross section of the $(e,e^{\prime
}p)-$ reaction in the Laboratory system has the following form\cite{Forest83}%
: 
\begin{equation}
\frac{d^{6}\sigma }{d\Omega _{e}\text{ }d\Omega _{p}\text{ }d\varepsilon
_{2}dE_{p}\text{ }}=p_{p}E_{p}\sigma _{ep}S(E_{m},\overrightarrow{p_{m}}),
\label{2}
\end{equation}
where
\begin{equation}
\sigma _{ep}=\sigma _{mott}\left(
V_{C}W_{C}+V_{T}W_{T}+V_{I}W_{I}+V_{S}W_{S}\right)  \label{Eq. 3}
\end{equation}
is the off-shell \ electron-nucleon cross section, $S(E_{m},%
\overrightarrow{p_{m}})$ \ is the spectral function which defines the
combined probability to find a bound proton with momentum $\overrightarrow{%
p_{m}}$ \ on the shell with separation energy \bigskip 
$E_{m}.$

The kinematic functions $V$ in Eq. (\ref{Eq. 3})\ \ can be expressed ,
neglecting the mass of the electron, as:

\begin{eqnarray}
V_{C} &=&\frac{q_{_{\mu }}^{4}}{q^{4}}, \\
V_{T} &=&\frac{q_{_{\mu }}^{2}}{2q^{2}}+\tan ^{2}(\frac{\theta _{e}}{2}), \\
V_{I} &=&\frac{q_{_{\mu }}^{2}}{q^{2}}\cos \phi \sqrt{\frac{q_{_{\mu }}^{2}}{%
q^{2}}+\tan ^{2}(\frac{\theta _{e}}{2})}, \\
V_{S} &=&\frac{q_{_{\mu }}^{2}}{q^{2}}\cos ^{2}\phi +\tan ^{2}(\frac{\theta
_{e}}{2}),
\end{eqnarray}

\noindent
and 
\begin{equation}
\sigma _{mott}=\frac{\alpha ^{2}\cos ^{2}\frac{\theta _{e}}{2}}{4\varepsilon
_{1}^{2}\sin ^{4}\frac{\theta _{e}}{2}}(1+\frac{2\varepsilon _{1}}{m_{p}}%
\sin ^{2}\frac{\theta _{e}}{2})^{-1}
\end{equation}

\noindent
is the Mott cross section, $\ \phi $ \ is the angle between the scattering
plane and the plane defined by the vectors \ $\overrightarrow{p}_{p}$ and $\ 
\overrightarrow{q}.$

For the structure functions W in Eq.(\ref{Eq. 3}) we use the off-shell
prescription of de Forest \cite{Forest83}:

\begin{eqnarray}
W_{C} &=&\frac{1}{4\bar{E}E_{p}}\{(\bar{E}+E_{p})^{2}(F_{1}^{2}+\frac{%
\overline{q}_{\mu }^{2}}{4m_{p}^{2}}\kappa
_{p}^{2}F_{2}^{2})-q^{2}(F_{1}+\kappa _{p}F_{2})^{2}\},  \label{211} \\
W_{T} &=&\frac{\overline{q}_{\mu }^{2}}{2\bar{E}E_{p}}(F_{1}+\kappa
_{p}F_{2})^{2},  \nonumber \\
W_{I} &=&-\frac{p_{p}\sin \gamma }{\bar{E}E_{p}}(\bar{E}+E_{p})(F_{1}^{2}+%
\frac{\overline{q}_{\mu }^{2}}{4m_{p}^{2}}\kappa _{p}^{2}F_{2}^{2}), 
\nonumber \\
W_{S} &=&\frac{p_{p}^{2}\sin ^{2}\gamma }{\bar{E}E_{p}}(F_{1}^{2}+\frac{%
\overline{q}_{\mu }^{2}}{4m_{p}^{2}}\kappa _{p}^{2}F_{2}^{2}),  \nonumber
\end{eqnarray}
where $\kappa _{p\text{ }}$=1.793 is the anomalous magnetic moment of the
proton \ in units of the Bohr magneton ,

\begin{equation}
\bar{E}=\sqrt{p_{m}^{2}+m_{p}^{2}}\text{,}
\end{equation}
$m_{p}$ is the mass of the proton, $\overline{q}_{\mu }$ $=(\overrightarrow{q%
},i\overline{\omega })\ ,\ \overline{\omega }=E_{p}-\ \ \overline{E},$ $%
\gamma $\ \ is the angle between \ $\overrightarrow{p}_{p}$ and $\ 
\overrightarrow{q},$


$F_{1}$ and $F_{2}$ are the on-shell Dirac and Pauli proton form factors,
respectively:

\begin{equation}
F_{1}(q_{_{\mu }}^{2})=\frac{1}{1+\frac{q_{_{\mu }}^{2}}{4m_{p}^{2}}}%
[G_{E}(q_{_{\mu }}^{2})+\frac{q_{_{\mu }}^{2}}{4m_{p}^{2}}G_{M}(q_{_{\mu
}}^{2})],
\end{equation}

\begin{equation}
\kappa _{p}F_{2}(q)=\frac{1}{1+\frac{q_{_{\mu }}^{2}}{4m_{p}^{2}}}%
[G_{M}(q_{_{\mu }}^{2})-G_{E}(q_{_{\mu }}^{2})],
\end{equation}

\bigskip

\bigskip
\noindent 
where 
\begin{equation}
G_{E}(q_{_{\mu }}^{2})=(\frac{1}{1+\frac{q_{_{\mu }}^{2}}{0.71}})^{-2},
\label{Eqn 13}
\end{equation}

\begin{equation}
G_{M}(q_{_{\mu }}^{2})=\mu _{p}G_{E}(q_{_{\mu }}^{2})],
\end{equation}
$\mu _{p}=2.793$ is the proton magnetic moment \ in units of the Bohr
magneton and $q_{_{\mu }}^{2}$ in Eq.(\ref{Eqn 13}) is in \ (GeV/c)$^{2}.$

In the independent particle shell model the spectral function for the
spherical orbitals $\alpha \equiv nlj$ with binding energy $E_{\alpha }$
takes the simple form:

\begin{equation}
S(E_{m},\overrightarrow{p_{m}})=\delta \left( E-E_{\alpha }\right) \text{ }%
\upsilon _{\alpha }^{2}\text{ }n_{\alpha }\left( \overrightarrow{p_{m}}%
\right) ,
\end{equation}
where $\upsilon _{\alpha }^{2}$ and $n_{\alpha }\left( p\right) $ are the
occupation number and momentum distribution of the $\alpha $ orbital,
respectively. The six folded $(e,e^{\prime }p)-$cross section could be
transformed into a five folded one:

\begin{equation}
\frac{d^{5}\sigma }{d\Omega _{e}\text{ }d\Omega _{p}\text{ }dE_{p}\text{ }}%
=p_{p}E_{p}\sigma _{ep}\ \upsilon _{\alpha }^{2}n_{\alpha }\left( 
\overrightarrow{p_{m}}\right) ,
\end{equation}
where it is imposed energy and momentum conservation for the kinematics
variables \ in $\sigma _{ep}$.

\section{SINGLE PARTICLE BOUND STATES}

The single particle bound state energies and momentum distributions were
calculated in the framework of the macroscopic-microscopic approach by using
the BARRIER code \cite{Barrier99}.

The energy of the nucleus is presented as:

\begin{equation}
E_{tot}=E_{LD}+\delta E_{shell},  \label{111}
\end{equation}
where $E_{LD}$ $\ \ $is the macroscopic liquid drop \ part of the energy and 
$\delta E_{shell}$\ \ is\ the shell correction, which describes shell and
pairing effects. Both shell correction and the macroscopic part of the
energy have been calculated according to \cite{Barrier99}.

\subsection{Nuclear shape parametrization}

Only axially symmetric nuclear shapes have been considered in the present
work, and the deformed shape (up to and beyond its separation into two
fragments) can be conveniently described by the Cassini ovoids \cite
{Pashkevich71,Ivanyuk97}. The potential-energy surfaces are calculated as
functions of $\varepsilon $ (elongation) and ${\alpha _{4}}$ (hexadecapolar
momentum). From these potential energy surfaces, the equilibrium (ground
state) deformation parameters ${\varepsilon }$ and ${\alpha _{4}}$ were
calculated by minimizing the total nuclear energy ( Eq.(\ref{111})): $%
\varepsilon =0.227$ and $\alpha _{4}=0.059$.

\subsection{Nuclear Potential}

An Woods- Saxon potential\cite{WoSa99}, consisting of the central part $V$,
spin-orbit $V_{SO}$, and the Coulomb potential $V_{Coul}$ for protons, was
employed:

\begin{equation}
V^{WS}(r,z,\varepsilon ,\widehat{\alpha })=V(r,z,\varepsilon ,\widehat{%
\alpha })+V_{so}(r,z,\varepsilon ,\widehat{\alpha })+V_{Coul}(r,z,%
\varepsilon ,\widehat{\alpha })
\end{equation}
The real potential $V(r,z,\varepsilon ,\widehat{\alpha })$ involves the
parameters V$_{0}$, $r_{0}$ and $a$ , describing the depth of the central
potential, the radius and the diffuseness parameter, respectively, and it is
expressed as:

\begin{equation}
V(r,z,\varepsilon ,\widehat{\alpha })=\frac{V_{0}}{1+exp\left[ \frac{%
dist(r,z,\varepsilon ,\widehat{\alpha })}{a}\right] },
\end{equation}
where $dist(r,z,\varepsilon ,\widehat{\alpha })$ is the distance between a
point and the nuclear surface, and $\varepsilon $ and $\widehat{\alpha }$
are deformation parameters.

The depth of the central potential is parametrized as

\begin{equation}
V_{0}=V_{0}[1\pm \kappa (N-Z)/(N+Z)],
\end{equation}
with the plus sign for protons and the minus sign for neutrons, with the
constant $\kappa =0.63$.

The spin-orbit interaction is then given by:

\begin{equation}
V_{so}(r,z,\varepsilon ,\widehat{\alpha })=\lambda \left( \frac{h}{2Mc}%
\right) ^{2}\nabla V(r,z,\varepsilon ,\widehat{\alpha })\cdot (\vec{\sigma}%
\times \vec{p}),
\end{equation}
where $\lambda $ denotes the strength of the spin--orbit potential and M is
the nucleon mass. The vector operator $\vec{\sigma}$ stands for Pauli
matrices and $\vec{p}$ is the linear momentum operator.

The Coulomb potential is assumed to be that corresponding to the nuclear
charge $(Z-1)e$, taken to be uniformly distributed inside the nucleus. It is
computed in cylindrical coordinates by using the expression given in \cite
{Pashkevich71}.

\subsection{Single particle potential parameter definitions}

For the ground state deformation of $^{238}$U, small changes in $\lambda $
(spin-orbit potential strength) and $r_{0-so}$ (spin-orbit potential radius)
of the Chepurnov parameters\cite{Chepurnov68} are introduced in order to
reproduce adequately the spin/parity of the levels sequence. Using single
particle states obtained by this procedure, the quasiparticle states can be
calculated for the first minimum region, providing spin, parity, energy and
level spacing for the ground and some low--lying states. The quasiparticle
spectrum was obtained by using the semi-microscopic combined method \cite
{Dencom95}.

The potential parameters were chosen to give the best fit to the spectrum of
single-quasiparticle excitations of the Z-odd neighboring nuclei $^{239}$Np.

\subsection{Single particle wave functions}

The Hamiltonian matrix elements are calculated with the wave functions of a
deformed axially symmetric oscillator\ potential. The wave functions in the
coordinate space $\phi _{i}$ are expanded into eigenfunctions of the axially
deformed harmonic oscillator potential.

\bigskip

These eigenfunctions form a complete orthonormal basis for the single
particle Woods-Saxon wave function

\begin{equation}
\Psi _{i}\left( \overrightarrow{R},\sigma \right) =\sum\limits_{n_{\rho
},n_{z},\Lambda ,\Sigma }C_{n_{\rho },n_{z},\Lambda ,\Sigma }^{i}\Phi
_{n_{\rho },n_{z},\Lambda ,\Sigma }\left( \overrightarrow{R},\sigma \right) .
\end{equation}

From this expansion, we may conveniently express the single particle
Woods-Saxon wave function in momentum space:

\begin{equation}
\widetilde{\Psi _{i}}\left( \overrightarrow{K},\sigma \right) =\sum_{n_{\rho
},n_{z},\Lambda ,\Sigma }C_{n_{\rho },n_{z},\Lambda ,\Sigma }^{i}\widetilde{%
\Phi }_{n_{\rho },n_{z},\Lambda ,\Sigma }\left( \overrightarrow{K},\sigma
\right) ,
\end{equation}
with

\begin{equation}
\widetilde{\Phi }_{n_{\rho },n_{z},\Lambda ,\Sigma }\left( \overrightarrow{R}%
,\sigma \right) =\frac{1}{2\pi ^{3/2}}\int d\overrightarrow{R}e^{-i\text{ }%
\overrightarrow{K}\cdot \overrightarrow{R}}\Phi _{n_{\rho },n_{z},\Lambda
,\Sigma }\left( \overrightarrow{R},\sigma \right)
\end{equation}
normalized to one.

\bigskip

We define densities $\ n_{i}\left( \overrightarrow{K}\right) $ in momentum
space \ in an analogous way of that in the configuration space:

\begin{equation}
\rho _{i}\left( \overrightarrow{R}\right) =\rho _{i}\left( r,z\right)
=\left| \Phi _{i}^{+}\left( r,z\right) \right| ^{2}+\left| \Phi
_{i}^{-}\left( r,z\right) \right| ^{2},
\end{equation}
with

\begin{equation}
\Phi _{i}^{\pm }\left( r,z\right) =\frac{1}{\sqrt{2\pi }}\sum_{n_{\rho
},n_{z},\Lambda ,\Sigma }\delta _{\Sigma ,\pm \frac{1}{2}}\delta _{\Lambda
,\pm \Lambda }C_{n_{\rho },n_{z},\Lambda ,\Sigma }^{i}\Phi _{n_{\rho
},n_{z},\Lambda ,\Sigma }\left( \overrightarrow{R},\sigma \right) ,
\end{equation}
and

\begin{equation}
n_{i}\left( \overrightarrow{K}\right) =n_{i}\left( k,k_{z}\right) =\left| 
\widetilde{\Phi }_{i}^{+}\left( k,k_{z}\right) \right| ^{2}+\left| 
\widetilde{\Phi }_{i}^{-}\left( k,k_{z}\right) \right| ^{2},
\end{equation}
with

\begin{equation}
\widetilde{\Phi }_{i}^{\pm }\left( k,k_{z}\right) =\frac{1}{\sqrt{2\pi }}%
\sum_{n_{\rho },n_{z},\Lambda ,\Sigma }\delta _{\Sigma ,\pm \frac{1}{2}%
}\delta _{\Lambda ,\pm \Lambda }C_{n_{\rho },n_{z},\Lambda ,\Sigma }^{i}%
\widetilde{\Phi }_{n_{\rho },n_{z},\Lambda ,\Sigma }\left( \overrightarrow{K}%
,\sigma \right) .
\end{equation}

These single particle momentum distributions $n_{i}\left( \overrightarrow{K}%
\right) $ were averaged over nuclear symmetry axis directions.

\bigskip

Similarly to the total density

\begin{equation}
\rho \left( \overrightarrow{R}\right) =\sum_{i}2\upsilon _{i}^{2}\rho
_{i}\left( \overrightarrow{R}\right) ,
\end{equation}
the total momentum distribution is given by

\begin{equation}
n\left( \overrightarrow{K}\right) =\sum_{i}2\upsilon _{i}^{2}n_{i}\left( 
\overrightarrow{K}\right) ,
\end{equation}
where $\upsilon _{i}^{2}$ is the occupation probability resulting from the
BCS model.\cite{Dencom93},\cite{Dencom95}

The results for the occupation number calculations are shown in Fig. 2.

The energies of the $^{238}U$ \ proton bound states are shown in table 1.

\bigskip

\section{FISSILITY}

\bigskip

\bigskip 
The quasifree knockout of nucleons leads to the excitation of the
residual nucleus. This excitation energy ($E^{\ast }$, nucleus A-1) has two
origins: holes in the shells of the nucleus A, which appear as a result of
the knockout of nucleons, and final state interaction (FSI) of the outgoing
nucleon, which we assume as negligeable \ due to the high energy of the
proton.\bigskip

The fast, quasi free \ reaction stage occurs \ at zero thermal excitation
(ground state) of the initial nucleus $^{238}U,$ and results in a single
hole in one of the shells. This single hole configuration \ \ forms a
doorway for a thermalization process which leads to the thermal excitation $%
E^{\ast }$of the residual nucleus $^{237}Pa.$

The thermalization is a complicate process which involves creation of new \
many particle-hole configurations in competition with particles emission and
fission, and for some doorway configurations it might has non statistical
character, but, as a first guide-line for order of magnitude estimates we \
calculate the total fission probability (nucleus with energy $E^{\ast }$
deexcites in several steps) on the statistical \ theory \ grounds, both \
with \ and without taking into account the preequilibrium \ decay.

\subsection{Compound nucleus model \ }

\bigskip 
Firstly, we considered a extreme situation, by assuming that the
residual interaction leads to thermalization and formation \ of compound
nucleus just after the fast reaction stage, without any preequilibrium
particle emission \ In this case, the compound nucleus excitation energy is
assumed to be :

\begin{equation}
\ E^{\ast }=-E_{\alpha }
\end{equation}

where $E_{\alpha }$ \ is the energy of the bound state (hole).

For calculations of compound nucleus fissility \ we used \ the Bohr Wheeler 
\cite{Bohr} and Weisskopf \cite{Weisskopf}\ models\ for the description of
the evaporation/fission competition. It was developed a Monte Carlo
algorithm for the evaporation/fission processes which includes not only the
neutron evaporation vs. fission competition, but also takes into account the
proton and alpha-particle contributions.

The probability for the emission of a particle $j$ with kinetic energy
between $E_{k}$ and $E_{k}+dE_{k}$ is calculated within the Weisskopf
statistical model\cite{Weisskopf} as:

\begin{equation}
P_{f}(E_{k})dE_{k}=\gamma _{j}\sigma _{j}E_{k}\left( \frac{\rho _{f}}{\rho
_{i}}\right) dE_{k},  \label{11}
\end{equation}
where $\sigma _{j}$ is the nuclear capture cross section for the particle $j$%
, $\gamma _{j}=\frac{gm}{\pi ^{2}h^{3}}$, where $g$ denotes the number of
spin states, and $m$ is the particle mass. The level densities for the
initial and final nucleus, $\rho _{i}$ and $\rho _{f}$, respectively, are
calculated from the Fermi gas expression

\[
\rho (E_{j}^{\ast })=\exp \left[ 2(\text{ }aE_{j}^{\ast })^{1/2}\right] , 
\]
where $a$ is the level density parameter \ (see below),

\begin{equation}
\ E_{j}^{\ast }=E^{\star }-\left( B_{j}+V_{j}\right) ,  \label{13}
\end{equation}
$E^{\ast }$ is the nuclear excitation energy in the initial state, $B_{j}$
is the particle separation energy, and $V_{j}$ is the Coulomb barrier
corrected for the nuclear temperature, $\tau $, defined by $E^{\ast }=a\tau
^{2}.$

The particle emission width is calculated as

\begin{equation}
\Gamma _{j}=\int_{0}^{\ E_{j}^{\ast }}P_{j}(E_{k})dE_{k}.  \label{14}
\end{equation}

From this general equation, the $\ k$-particle emission probability relative
to the $j$-particle emission is:

\begin{equation}
\frac{\Gamma _{k}}{\Gamma _{j}}=\left( \frac{\gamma _{k}}{\gamma _{j}}\frac{%
\ E_{k}^{\ast }}{\ E_{j}^{\ast }}\frac{a_{j}}{a_{k}}\right) \exp \left[
2\left( \left( a_{k}\ E_{k}^{\ast }\right) ^{1/2}-\left( a_{j}E_{j}^{\ast
}\right) ^{1/2}\right) \right] .  \label{15}
\end{equation}

\bigskip 

The level density parameter for neutron emission is \cite{Iljinov}:

\begin{equation}
a_{n}=(0.134\text{ }A-1.21\text{ }\cdot 10^{-4}\text{ }A^{2})\text{ MeV}%
^{-1},  \label{16}
\end{equation}
and for all other particle emission this quantity is related to $a_{n}$ by,

\begin{equation}
a_{j}=r_{j}a_{n},  \label{17}
\end{equation}
where $r_{j}$ is an adimensional constant.

Shell model corrections \cite{Guaraldo} are not taken into account. For high
excitation energies their effects are likely to cancel each other upon
averaging over all possible nuclei created during the reaction.

\bigskip 
Using the fission width from \ the liquid drop model \cite{Bohr},
and the neutron emission width from Weisskopt \cite{Weisskopf}, we get

\begin{equation}
\frac{\Gamma _{f}}{\Gamma _{n}}=K_{f}\text{ }\exp \left[ 2\left( \left(
a_{f}\ E_{f}^{\ast }\right) ^{1/2}-\left( a_{n}\ E_{n}^{\ast }\right)
^{1/2}\right) \right] ,  \label{18}
\end{equation}
where

\begin{equation}
K_{f}=K_{0}\text{ }a_{n}\frac{2((a_{f}\ E_{f}^{\ast })^{1/2}-1}{%
4A^{2/3}a_{f}E_{n}^{\ast }},  \label{19}
\end{equation}
with $K_{0}=14.39$ MeV and $E_{j}^{\ast }=E^{\ast }-B_{f}.$ Here $B_{f}$ is
the fission barrier height discussed below.

For proton emission we get

\begin{equation}
\frac{\Gamma _{p}}{\Gamma _{n}}=\left( \frac{\ E_{p}^{\ast }}{\ E_{n}^{\ast }%
}\right) \exp \left[ 2\left( a_{n}\right) ^{1/2}(\left( r_{p}E_{p}^{\ast
}\right) ^{1/2}-(E_{n}^{\ast })^{1/2})\right] ,  \label{20}
\end{equation}
and for alpha-particle emission\cite{Vandenbosch}\cite{Iljinov},

\begin{equation}
\frac{\Gamma _{a}}{\Gamma _{n}}=\left( \frac{2\ E_{a}^{\ast }}{\ E_{n}^{\ast
}}\right) \exp \left[ 2(a_{n}\ )^{1/2}((r_{a}E_{a}^{\ast
})^{1/2}-(E_{n}^{\ast })^{1/2})\right] .  \label{21}
\end{equation}

\bigskip 
In the above equations, the Coulomb potential for protons is \cite
{Tavares}

\begin{equation}
V_{p}=\ C\frac{k_{p}(Z-1)e^{2}}{r_{0}(A-4)^{1/3}+R_{p}},  \label{22}
\end{equation}
and for alphas,

\begin{equation}
V_{a}=C\frac{2K_{a}(Z-2)e^{2}}{r_{0}(A-4)^{1/3}+R_{a}},  \label{23}
\end{equation}
where $K_{p}$ $=0.70$ and $K_{a}$ $=0.83$ are the Coulomb barrier
penetrability for protons and alpha particles, respectively, $R_{p}$ $=1.14$
fm is the proton radius ,$R_{a}=2.16$ fm is the alpha particle radius, and $%
r_{0}=1.2$ $fm$ . The factor $C$ introduces in a semi-empirical way the
dynamical effects in particle separation energy and \ fission barrier due to
the nuclear temperature \cite{Tavares}, namely

\begin{equation}
C=1-\frac{E^{\ast }}{B}\ ,  \label{24}
\end{equation}
where $B$ is the total nuclear binding energy \ (B=1794 MeV for $^{237}Pa$ 
\cite{Tavares}).

The fission barrier is calculated by\cite{Tavares},

\begin{equation}
\ B_{f}=C(0.22(A-Z)-1.40Z+101.5)\text{ MeV.}  \label{25}
\end{equation}

\bigskip 
The neutron separation energy was taken as $5.78$ MeV for the first
step ($^{237}Pa$) , and for the other steps as\cite{Guaraldo}:

\begin{equation}
B_{n}=\ (-0.16(A-Z)+0.25Z+5.6)\text{ MeV,}  \label{26}
\end{equation}
while the proton and alpha-particle separation energies are calculated
through the nuclear mass formula\cite{Segre}:

\begin{equation}
B_{p}=m_{p}+M(A-1,Z-1)-M(A,Z),  \label{27}
\end{equation}
where $m_{p}$ is the proton mass, and $M(A,Z)$ is the nuclear mass
calculated with the parameters from reference \cite{Segre}. For the alpha
particles we get

\begin{equation}
B_{a}=m_{a}+M(A-4,Z-2)-M(A,Z),  \label{28}
\end{equation}
where $m_{a}$ is the alpha particle mass.

\bigskip 

These values reproduce the experimental data for P$_{f}$ (see
discussion below)

The present Monte Carlo code for evaporation-fission calculates, at each\
step $\ i$\ of the evaporation chain, the fission probability, $F_{i}$,
defined as 
\[
F_{i}=\frac{(\frac{\Gamma _{f}}{\Gamma _{n}})_{i}}{1+(\frac{\Gamma _{f}}{%
\Gamma _{n}})_{i}+(\frac{\Gamma _{p}}{\Gamma _{n}})_{i}+(\frac{\Gamma
_{\alpha }}{\Gamma _{n}})_{i}}, 
\]

An evaporating particle $\ j$ \ is randomly chosen (neutron, proton or alpha
particle), according to its relative branching ratios. Once one of these
particles is chosen , the mass and atomic numbers are recalculated through

\[
A_{i+1}=A_{i}-\Delta A_{i}, 
\]
and

\[
Z_{i+1}=Z_{i}-\Delta Z_{i}, 
\]
where $\Delta A_{i}$, and $\Delta Z_{i}$, are, respectively, the mass and \
\ atomic numbers of the ejected particle at the $ith$ step in the
evaporation process. The nuclear excitation energy is modified \ according
to the expression

\[
E_{i+1}^{\ast }=E_{i}^{\ast }-B_{i}-T_{i}, 
\]
where $B_{i}$\ and $T_{i}$\ are the separation and the asymptotic kinetic
energies of the particle being ejected, respectively. For neutrons $T=2$
MeV, and for protons and alpha particles $T=0$ \ MeV$.$ The expressions\
described above ensure that the nuclear excitation energy will be, at each
step in the evaporation chain, smaller than in the previous step. This
process continues until the excitation energy available in the nucleus is
not enough to emit any\ one of the possible evaporating particles. At this
point the evaporation process stops, and we can calculate the nuclear
fissility by the expression

\begin{equation}
W=\sum_{i}\left[ \prod_{j=0}^{i-1}(1-F_{j})\right] F_{i}.
\end{equation}

Using the model described above, we calculated the fissility for\ $^{237}Pa$
(figure 3, solid curve). Peaks observed for the fissility reflect the
opening of the fission channel in the daughter nuclei. Figure 3 also shows (
rectangle) the data for the fissility of $^{237}Pa$ obtained by
extrapolation of the neutron to fission widths ratios for $Z=91,$ and $%
A=230,231,232,233$ to $A=237$ \ \cite{Gavron}, by using the empirical trend
presented in Vandenbosch and Huizenga \cite{Vandenbosch}

It should be pointed out that in our calculations of \ the fissility we
assumed that the hole excitation energies for an A-1 nucleus correspond to
the compound nucleus excitation energies , that is to say, the complete
thermalization is reached without any preequilibrium decay. Such
calculations could be considered as an upper limit estimate for the
fissility.

\bigskip

\subsection{\ Exciton model}

\bigskip 
During the thermalization \ of\ the hole excitation energy, the
nucleus $A-1$ \ could undergo particle evaporation (preequilibrium decay 
\cite{Bonetti}, \cite{Gadioli}).

\bigskip
 In this case, the energy of the hole is not attributed to the
nuclear temperature but, instead, assumed as a characteristic of the doorway
state in the thermalization process \ followed by the emission of particles
or fission.

The calculation involving \ the preequilibrium decay was performed within
the framework of the exciton model{\ \cite{Kalbach}}, using the code STAPRE.
In this model, the states of the system are classified according to the
number of excitons $n$, which corresponds to the total number of excited
particle $p$ and hole $h$ degrees of freedom, $n=p+h$. The exciton model
included in STAPRE does not distinguish between protons and neutrons.
Starting from a simple configuration of low exciton number, the system is
assumed to equilibrate through a series of two-body collisions and to emit
particles from all intermediate states. The application of a two-body
interaction to the states of a $(p,h)$ configuration results in states with $%
(p+1,h+1)$, $(p,h)$, and $(p-1,h-1)$ excited particles and holes. The
difference between the number of excited particles and holes remains fixed,
justifying the use of the exciton number to label the states. However, the
transition rates, which are an averaging over all states of a configuration,
do depend on the number of excited particles and holes. The equation
governing the time development of the occupation $P(n)$ of the $n-th $
exciton configuration can thus be written as 
\begin{equation}
\frac{dP(n)}{dt}=\lambda _{-}(n+2)\,P(n+2)+\lambda _{0}(n)\,P(n)+\lambda
_{+}(n-2)\,P(n-2)-\lambda (n)\,P(n)  \label{preeqeq}
\end{equation}
where $\lambda (n)$ is the total transition rate, 
\begin{equation}
\lambda (n)=\lambda _{-}(n)+\lambda _{0}(n)+\lambda _{+}(n)+\lambda _{e}(n),
\label{55}
\end{equation}
with $\lambda _{e}(n)$ being the total rate of particle emission from the $%
n-th$ exciton configuration. The quantities $\lambda _{-}(n)$, $\lambda
_{0}(n)$, and $\lambda _{+}(n)$ are the average rates for internal
transitions from the $n-th$ exciton configuration with a change of exciton
numbers by -2, 0, or +2.

The internal transition rates can be written as the product of the average
squared matrix element of the residual interaction $|M|^{2}$with the
relative density of available states. For the latter, STAPRE uses the
expressions of Williams{\ \cite{Williams}} as corrected for the Pauli
principle by Cline{\ \cite{Cline2}}. These yield 
\begin{eqnarray}
\lambda _{-}(n)\equiv \lambda _{-}(p,h,E) &=&\frac{2\pi }{\hbar }|M|^{2}%
\frac{g\left( gE-C_{p+1,h+1}\right) ^{2}}{p+h+1},  \label{56} \\
\lambda _{\,0\,}(n)\equiv \lambda _{\,0\,}(p,h,E) &=&\frac{2\pi }{\hbar }%
|M|^{2}g(p+h-1)\left( gE-C_{p,h}\right) ,  \nonumber \\
\lambda _{+}(n)\equiv \lambda _{+}(p,h,E) &=&\frac{2\pi }{\hbar }%
|M|^{2}gph(p+h-2),  \nonumber
\end{eqnarray}
where 
\[
C_{p,h}=\frac{1}{2}\left( p^{2}+h^{2}\right) , 
\]
with E the excitation energy of the system. The parameter g is the
single-particle state density, which is taken to be $g=\frac{6}{\pi ^{2}}a$,
with $a$ \ as the level density parameter. Following Kalbach-Cline{\cite
{Kalbach-Cline}}, the average matrix element is approximated as 
\[
|M|^{2}=\frac{f_{M}}{A^{3}E}, 
\]
where $A$ is the mass number of the system and $f_{M}$ is a parameter, which
we assumed to be $f_{M}=230$ MeV$^{3}$ in our calculations.

The particle emission rate $\lambda _{e}(n)$ is the sum of the integrated
proton and neutron differential emission rates, $\lambda _{e\nu
}(n,\varepsilon )\,d\varepsilon _{\nu }$, which are determined through
considerations of detailed balance\cite{Cline3}, 
\begin{eqnarray}
\lambda _{e\nu }(n,\varepsilon )\,d\varepsilon _{\nu } &\equiv &\lambda
_{e\nu }(p,h,E,\varepsilon _{\nu })\,d\varepsilon _{\nu }  \label{57} \\
&=&\frac{1}{\pi ^{2}\hbar ^{3}}\mu _{\nu }\,\varepsilon _{\nu }\,\sigma
_{\nu }(\varepsilon _{\nu })\,R_{\nu }\frac{\omega (p-1,h,E-B_{\nu
}-\varepsilon _{\nu })}{\omega (p,h,E)}\,d\varepsilon _{\nu },  \nonumber
\end{eqnarray}
where $\mu _{\nu }$ is the reduced mass of the emitted neutron/proton, $%
\varepsilon _{\nu }$ its outgoing kinetic energy, $B_{\nu }$ its separation
energy, and $\sigma _{\nu }(\varepsilon _{\nu })$ is the cross section for
the inverse absorption process. The factor $R_{\nu }$ is a simple correction
standing for the fact that neutrons and protons have not been distinguished
in the process; thus, 
\[
R_{\nu }=\left\{ 
\begin{array}{c}
N/A\qquad \mbox{for neutron emission} \\ 
Z/A\qquad \mbox{for proton emission}
\end{array}
.\right. 
\]
The densities of states are taken to be the Williams densities, 
\begin{equation}
\omega (p,h,E)=\frac{g\,\left( gE-A_{p,h}\right) ^{p+h-1}}{p!\,h!\,(p+h-1)!},
\label{58}
\end{equation}
where the Pauli blocking correction is 
\[
A_{p,h}=\frac{1}{4}\left( p^{2}+h^{2}+p-3h\right) . 
\]
The differential emission rates differ from those of usual Weisskopf
compound nucleus emission by the factor $R_{\nu }$ and by the use of exciton
state densities rather than compound nucleus ones.

The time evolution equation, Eq. (\ref{preeqeq}), form a set of coupled
linear differential equations whose solution could be written in the form of
a vector as, 
\[
\vec{P}(t)=\exp \left[ -\Lambda \,t\right] \,\vec{P}_{0,} 
\]
where the matrix $\Lambda $ is given by 
\[
\Lambda _{nn^{\prime }}=\lambda (n)\,\delta _{n^{\prime },n}-\lambda
_{-}(n+2)\,\delta _{n^{\prime },n+2}-\lambda _{0}(n)\,\delta _{n^{\prime
},n}-\lambda _{+}(n-2)\,\delta _{n^{\prime },n-2}, 
\]
and the vector $\vec{P}_{0}$ describes the initial exciton configuration of
the system, 
\[
P_{0}(n)\equiv P_{0}(p,h)=\delta _{p,p_{0}}\delta _{h,h_{0}}. 
\]
The decay of the system into all possible final configurations can be
obtained by integrating the total emission rate over all time, 
\begin{equation}
\sum_{n}\,\int_{0}^{\infty }\lambda _{e}(n)\,P(n,t)\,dt=\sum_{n,n^{\prime
}}\lambda _{e}(n)\,\left( \Lambda ^{-1}\right) _{n,n^{\prime
}}P_{0}(n^{\prime }).  \label{59}
\end{equation}
The decay of the fraction of the initial probability, which survives
preequilibrium emission, is described using the Hauser-Feshbach formalism.
We have considered fission in competition with neutron and gamma emission.

The initial configuration in $^{237}$Pa \ consists of one-particle at the
Fermi level and one-hole in a bound state. This configuration is consistent
with the proton knockout reaction for $^{238}U$ \ initiating the statistical
cascade. Our calculations were performed \ assuming an one -hole initial
configuration \ of \ the l=0 partial wave alone. The particle at the \ Fermi
level contributes negligibly to the equilibration process. The fission
barriers, neutron separation energies \ and level density parameters were
taken the same as those of the \ compound nucleus calculations in the
previous section.

The exciton model fissility results for single hole states of $^{237}$Pa \
are shown in fig 3 by the dotted curve. We note \ that these calculations \
for fissility show a \ smoother behavior than that for compound model. The
preequilibrium \ particle emission \ removes \ some excitation energy \
before a equilibrium is reached reducing, therefore, the probability of
opening new \ chances for fission.

\bigskip

\section{Final results}

The differential cross section for the\ $(e,e^{\prime }pf)$-reaction was
obtained \ \ by assuming an isotropic angular distribution for the fission
fragments, and the fissility as a factor :

\bigskip 
\begin{equation}
\frac{d^{7}\sigma }{d\Omega _{e}\text{ }d\Omega _{p}\text{ }dE_{p}\text{ }%
d\Omega _{f}}=\frac{1}{4\pi }\frac{d^{5}\sigma }{d\Omega _{e}\text{ }d\Omega
_{p}\text{ }dE_{p}\text{ }}P_{f}  \label{60}
\end{equation}

\bigskip

Fig. 4 shows the seven folded differential cross sections$\frac{d^{7}\sigma 
}{d\Omega _{e}\text{ }d\Omega _{p}\text{ }dE_{p}\text{ }d\Omega _{f}}$ for
some bound proton states\ (table 1) and the compound nucleus model fissility
(solid curve in fig.3) calculated \ for $\varepsilon _{1}$=2000 MeV , $%
\theta e=23^{0}$ and the parallel kinematics \cite{Potter}. In this
kinematics \ $\varepsilon _{1}$ and $\theta e$ are fixed, \ and for each
value of $\varepsilon _{2}$ the proton spectrum \ is measured in the
direction of $\overrightarrow{q}$ ,\ varying \ each time the angle $\theta
_{P_{p}k_{1}}.$(see fig.1). \ For such scheme of measurements \ the initial
(missing ) momentum of the proton $\overrightarrow{p_{m}}$ \ is always
parallel (or antiparallel) $\ $to $\overrightarrow{q}$ . The parallel
kinematics simplify the accounting of FSI, \ since there are no contribution
of interference terms in the cross sections (see Eqs.\ (\ref{211})). Figs 5
and 6\ show the momentum distributions for states used in the calculation of
the cross sections presented in fig. 4, and fig. 7 shows the outgoing proton
kinetic energy and angle $\theta _{Ppk_{1}}$ versus \ $\varepsilon _{2}$ for
the parallel \ kinematics we use.

Figs. 8 and 9 show the differential \ $(e,e^{\prime }pf)-$cross sections
calculated for the same $\varepsilon _{1}$ and $\theta e$ but for two fixed
\ proton angles: $\theta _{P_{p}k_{1}}=0.98$ rad for the group of proton
states of $^{238}U$ which have a maximum in the low missing momentum region\
(fig.5), and $\theta _{P_{p}k_{1}}=\ 0.82$ rad for the group having a
maximum in the high missing momentum region\ (fig.6) . These angles $\theta
_{P_{p}k_{1}}$were chosen in order to achieve parallel kinematics, that is $%
\theta _{P_{p}q}\thickapprox 0$ ,\ and maximum for cross sections at both
the low ( $\theta _{P}=0.98$ rad ) and high ($\theta _{P}=0.82$ rad\ )
missing momentum regions. Fig.10 shows the missing momentum P$_{m}$ and the
angle $\theta _{P_{p}q}$ as functions of the outgoing proton kinetic energy
for $\theta _{p}=0.98$ rad ( solid curve) and $0.83$ rad ( dashed curve).

It is seen from the figures 8,9\ and 10 that for such \ a choice the cross
sections have maxima \ at \ $E_{p}$ around $300$ and $400$ MeV and,\ for
these energies, the proton angles $\theta _{P_{p}q\text{ }}$are small
(parallel kinematics).

The differential cross-sections presented in figures 4, 8 and 9 correspond
to the situation when the hole excitation energies for an A-1 nucleus is the
compound nucleus excitation energies , that is to say, the complete
thermalization is reached without any preequilibrium decay. Such
calculations could be considered as an upper limit \ estimate for the cross
section

\bigskip

\section{Conclusions}

We presented a theoretical study for the quasifree electrofission of $%
^{238}U.$

The proton bound states were calculated in the framework of the
Macroscopic-Microscopic approach, using the axially deformed Woods-Saxon
single particle potential. The occupation numbers were calculated in the BCS
approach.

The exclusive differential cross sections for the quasi free scattering
reaction stage were calculated in PWIA using off-shell electron-nucleon
cross sections.

The fissility for the single hole states of the residual nucleus $^{237}Pa$
was calculated in the framework of two approaches: compound nucleus model
without taking into account the preequilibrium emission of the particles,
and the exciton model accounting for preequilibrium emission. Both models
exhibit \ the same general trend, but the fissility \ as given by \ the
preequilibrium model is smoother.

The obtained results could serve as a first guide-line on order of magnitude
estimates of the expected cross sections for quasi free electrofission of $%
^{238}U.$

\vspace{1cm}
\noindent
{\large \bf Acknowledgments}

The authors thank the Brazilian agencies CNPq and FAPESP for the partial
support to this work and the graduate students W.R. Carvalho, M.S. Vaudeluci
and L.F.R. Macedo for their help.

\newpage
\bigskip
\noindent
{\large \bf Table I:}\ Proton single-particle levels of $\ ^{238}$U. The
Fermi level is the level 46.

\[
\text{%
\begin{tabular}{|l|l|l|l|l|l|l|l|l|l|l|l|}
\hline
{\small \ } & {\small \ \ [MeV]} & {\small \ \ }$\pi ${\small J} & {\small \
\ \ \ \ [N n}$_{z}${\small \ }$\Lambda ${\small ]} &  & {\small \ [MeV]} & 
{\small \ \ }$\pi ${\small J} & {\small \ \ \ \ \ [N n}$_{z}${\small \ }$%
\Lambda ${\small ]} &  & {\small [MeV]} & {\small \ \ }$\pi ${\small J} & 
{\small \ \ \ \ \ [N n}$_{z}${\small \ }$\Lambda ${\small ]} \\ \hline
{\small 1} & {\small -33.685} & {\small \ 1/2 } & {\small 1/2 [ 0 0 0]} & 
{\small 23} & {\small -16.192 } & {\small -3/2 } & {\small 3/2 [ 3 0 1]} & 
{\small 45} & {\small -7.491 } & {\small \ 3/2 } & {\small 3/2 [ 4 0 2]} \\ 
\hline
{\small 2} & {\small -31.397 } & {\small -1/2 } & {\small 1/2 [ 1 1 0]} & 
{\small 24} & {\small -15.490 } & {\small -1/2 } & {\small 1/2 [ 3 0 1]} & 
{\small 46} & {\small -7.195 } & {\small \ 1/2 } & {\small 1/2 [ 4 0 0]} \\ 
\hline
{\small 3} & {\small -30.043 } & {\small -3/2 } & {\small 3/2 [ 1 0 1]} & 
{\small 25} & {\small -15.415 } & {\small \ \ 7/2 } & {\small 7/2 [ 4 1 3]}
& {\small 47} & {\small -6.277 } & {\small \ 5/2 } & {\small 5/2 [ 6 4 2]}
\\ \hline
{\small 4} & {\small -29.670 } & {\small -1/2 } & {\small 1/2 [ 1 0 1]} & 
{\small 26} & {\small -14.529 } & {\small \ \ 9/2 } & {\small 9/2 [ 4 0 4]}
& {\small 48} & {\small -6.189 } & {\small -5/2 } & {\small 5/2 [ 5 2 3]} \\ 
\hline
{\small 5} & {\small -28.141 } & {\small \ \ 1/2 } & {\small 1/2 [ 2 2 0]} & 
{\small 27} & {\small -14.302 } & {\small \ \ 3/2 } & {\small 3/2 [ 4 2 2]}
& {\small 49} & {\small -5.348 } & {\small -3/2 } & {\small 3/2 [ 5 2 1]} \\ 
\hline
{\small 6} & {\small -26.630 } & {\small \ \ 3/2 } & {\small 3/2 [ 2 1 1]} & 
{\small 28} & {\small -13.984 } & {\small -1/2 } & {\small 1/2 [ 5 3 0]} & 
{\small 50} & {\small -4.827 } & {\small \ 7/2 } & {\small 7/2 [ 6 3 3]} \\ 
\hline
{\small 7} & {\small -25.963 } & {\small \ \ 1/2 } & {\small 1/2 [ 2 1 1]} & 
{\small 29} & {\small -13.111 } & {\small \ \ 1/2 } & {\small 1/2 [ 4 2 0]}
& {\small 51} & {\small -4.340 } & {\small -7/2 } & {\small 7/2 [ 5 1 4]} \\ 
\hline
{\small 8} & {\small -25.542 } & {\small \ \ 5/2 } & {\small 5/2 [ 2 0 2]} & 
{\small 30} & {\small -13.091 } & {\small -3/2 } & {\small 3/2 [ 5 4 1]} & 
{\small 52} & {\small \ -3.949 } & {\small -1/2 } & {\small 1/2 [ 5 2 1]} \\ 
\hline
{\small 9} & {\small -24.473 } & {\small \ \ 3/2 } & {\small 3/2 [ 2 0 2]} & 
{\small 31} & {\small -12.383 } & {\small \ \ 5/2 } & {\small 5/2 [ 4 1 3]}
& {\small 53} & {\small -3.667 } & {\small \ \ 1/2 } & {\small 1/2 [ 6 5 1]}
\\ \hline
{\small 10} & {\small -24.025 } & {\small \ -1/2 } & {\small 1/2 [ 3 3 0]} & 
{\small 32} & {\small -11.735 } & {\small \ -5/2 } & {\small 5/2 [ 5 3 2]} & 
{\small 54} & {\small -3.465 } & {\small -5/2 } & {\small 5/2 [ 5 1 2]} \\ 
\hline
{\small 11} & {\small -22.836 } & {\small \ \ 1/2 } & {\small 1/2 [ 2 0 0]}
& {\small 33} & {\small -11.053 } & {\small \ \ 7/2 } & {\small 7/2 [ 4 0 4]}
& {\small 55} & {\small -3.417 } & {\small \ 9/2 } & {\small 9/2 [ 6 2 4]}
\\ \hline
{\small 12} & {\small -22.716} & {\small \ -3/2} & {\small 3/2 [ 3 2 1]} & 
{\small 34} & {\small -10.831 } & {\small \ \ 3/2 } & {\small 3/2 [ 4 1 1]}
& {\small 56} & {\small -3.051 } & {\small -9/2 } & {\small 9/2 [ 5 0 5]} \\ 
\hline
{\small 13} & {\small -21.614 } & {\small -1/2 } & {\small 1/2 [ 3 2 1]} & 
{\small 35} & {\small -10.388 } & {\small -1/2 } & {\small 1/2 [ 5 4 1]} & 
{\small 57} & {\small -2.261 } & {\small 11/2 } & {\small 11/2 [ 6 1 5]} \\ 
\hline
{\small 14} & {\small -21.292 } & {\small -5/2 } & {\small 5/2 [ 3 1 2]} & 
{\small 36} & {\small -10.280 } & {\small -7/2 } & {\small 7/2 [ 5 2 3]} & 
{\small 58} & {\small -2.207 } & {\small -1/2 } & {\small 1/2 [ 7 5 0]} \\ 
\hline
{\small 15} & {\small -20.333 } & {\small -7/2 } & {\small 7/2 [ 3 0 3]} & 
{\small 37} & {\small -9.794 } & {\small \ \ 1/2 } & {\small 1/2 [ 4 1 1]} & 
{\small 59} & {\small -2.182 } & {\small \ 7/2 } & {\small 7/2 [ 5 0 3]} \\ 
\hline
{\small 16} & {\small -19.621 } & {\small -3/2 } & {\small 3/2 [ 3 1 2]} & 
{\small 38} & {\small -9.301 } & {\small \ \ 5/2 } & {\small 5/2 [ 4 0 2]} & 
{\small 60} & {\small -1.773 } & {\small \ 3/2 } & {\small 3/2 [ 6 4 2]} \\ 
\hline
{\small 17} & {\small -19.254 } & {\small \ 1/2 } & {\small 1/2 [ 4 2 0]} & 
{\small 39} & {\small -9.054 } & {\small -9/2 } & {\small 9/2 [ 5 1 4]} & 
{\small 61} & {\small -1.669 } & {\small \ 1/2 } & {\small 1/2 [ 6 4 0]} \\ 
\hline
{\small 18} & {\small -18.229} & {\small -5/2 } & {\small 5/2 [ 3 0 3]} & 
{\small 40} & {\small -8.356 } & {\small -3/2 } & {\small 3/2 [ 5 3 2]} & 
{\small 62} & {\small -1.553 } & {\small -3/2 } & {\small 3/2 [ 7 4 1]} \\ 
\hline
{\small 19} & {\small -18.189 } & {\small \ \ 3/2 } & {\small 3/2 [ 4 3 1]}
& {\small 41} & {\small -8.276 } & {\small \ \ 1/2 } & {\small 1/2 [ 6 4 0]}
& {\small 63} & {\small -1.459 } & {\small 13/2 } & {\small 13/2 [ 6 0 6]}
\\ \hline
{\small 20} & {\small -18.130 } & {\small -1/2 } & {\small 1/2 [ 3 1 0]} & 
{\small 42} & {\small -8.217 } & {\small -11/2 } & {\small 11/2 [ 5 0 5]} & 
{\small 64} & {\small -1.118 } & {\small -3/2 } & {\small 3/2 [ 5 1 2]} \\ 
\hline
{\small 21} & {\small -16.730 } & {\small \ \ 5/2 } & {\small 5/2 [ 4 2 2]}
& {\small 43} & {\small -7.624 } & {\small -1/2 } & {\small 1/2 [ 5 3 0]} & 
{\small 65} & {\small -1.065 } & {\small -1/2 } & {\small 1/2 [ 5 1 0]} \\ 
\hline
{\small 22} & {\small -16.416 } & {\small \ \ 1/2 } & {\small 1/2 [ 4 3 1]}
& {\small 44} & {\small -7.597} & {\small \ 3/2 } & {\small 3/2 [ 6 5 1]} & 
{\small 66} & {\small -0.393 } & {\small -5/2 } & {\small 5/2 [ 7 5 2]} \\ 
\hline
\end{tabular}
} 
\]

\newpage

\noindent
{\large \bf Figure Captions}

\noindent{\bf Fig.1}. 
Momentum diagram of the $\ A(e,e^{\prime }p)A-1$ \ reaction in the
impulse approximation:
$\overrightarrow{k}_{1},$ and $\overrightarrow{k}_{2}$ are the initial and
final electron momenta, respectively; $\overrightarrow{p}_{m}$ is the
momentum of proton before interaction;$\ \ \overrightarrow{\text{ }p_{p}}$ \
is the momentum of knocked-out \ proton.

\noindent{\bf Fig.2}. Occupation probabilities for the single particle bound states of $\
^{238}$U.

\noindent{\bf Fig.3}. Fissility of $^{237}Pa$ \ vs the hole excitation energy.
The solid curve shows the compound nucleus model calculation, assuming that
the hole excitation energies correspond to the compound nucleus
excitation energies: the complete thermalization is reached without any
preequilibrium decay. The dotted curve corresponds to the exciton model
calculations, which take into account the preequilibrium decay. The
rectangle shows the extrapolated experimental data (se text for details).

\noindent{\bf Fig. 4}. Seven folded differential cross sections for\ the parallel
kinematics. \ The calculations of the cross sections were accomplished for $%
\varepsilon _{1}=2000$ MeV, $\theta e=23^{0},\phi =0^{0}$ . The cross section for \ E=-10.388 MeV\
\ state ( subbarier) is multiplied by 100.

\noindent{\bf Fig.5}. \ Momentum distributions for some proton bound states having maxima
at the low missing momentum region.\ 

\noindent{\bf Fig.6.} \ Momentum distributions for some proton bound states having maxima
at the high missing momentum region.\ 

\noindent{\bf Fig. 7}. Variation of the angle $\theta _{p_{p}k_{1}}$ \ (see fig.1) and \
the outgoing proton kinetic energy \ E$_{p}$ vs \ $\varepsilon _{2}$ for\
the parallel kinematics and $\ \varepsilon _{1}=2000$ MeV, \ $\theta
e=23^{0}.$

\noindent{\bf Fig.8}. Differential cross sections for some bound states of $\ ^{238}$U,
having maxima of the momentum distributions at the low missing momentum
region.\ The calculations of the cross sections were accomplished for $%
\varepsilon _{1}=2000$ MeV, $\theta e=23^{0},\phi =0^{0}$ and $\ \theta
_{p_{p}k_{1}}=0.98$ rad.

\noindent{\bf Fig.9}. The same as in figure.5, but for some bound states of $\ ^{238}$U,
having maxima of the momentum distributions at the high missing momentum
region\ and \ for $\theta _{p_{p}k_{1}}=0.82$ rad\ .

\noindent{\bf Fig.10}. Missing momentum p$_{m}$ and angle $\theta _{P_{p}q}$ as functions
of the outgoing proton kinetic energy for $\theta _{p_{p}k_{1}}=0.98$ rad (
solid curve) and $0.83$ rad ( dashed curve).

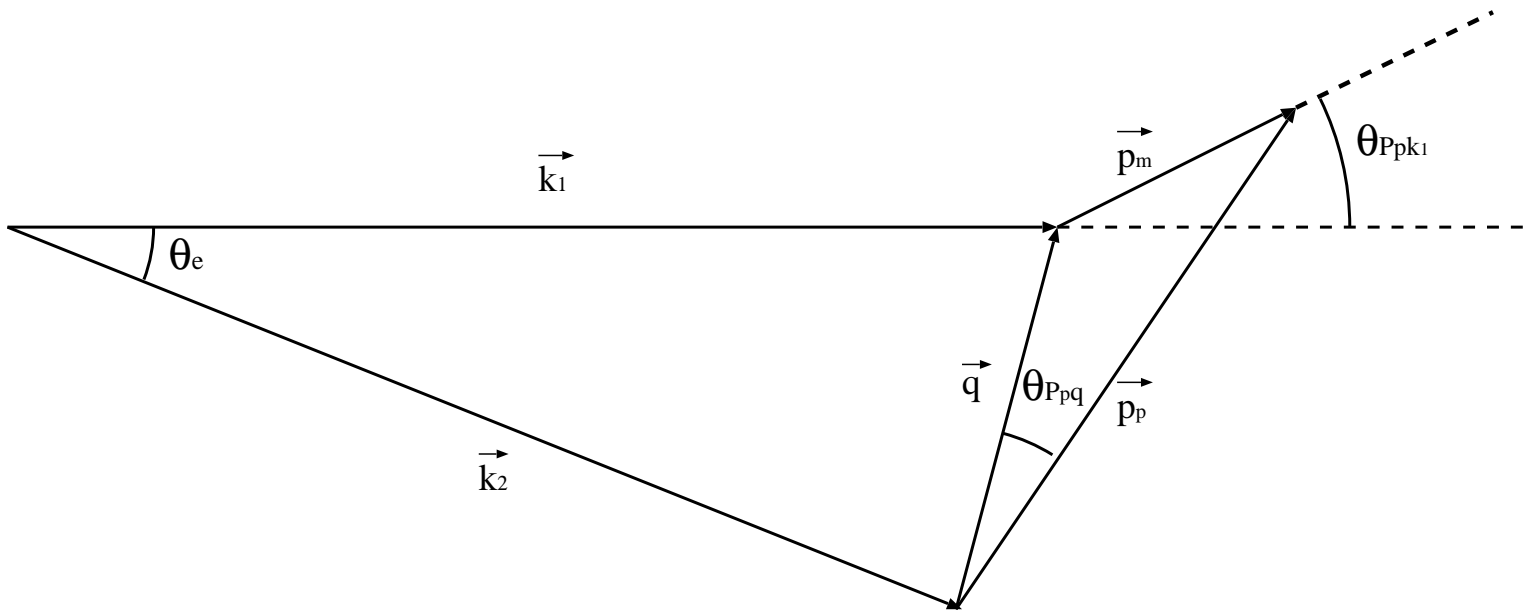
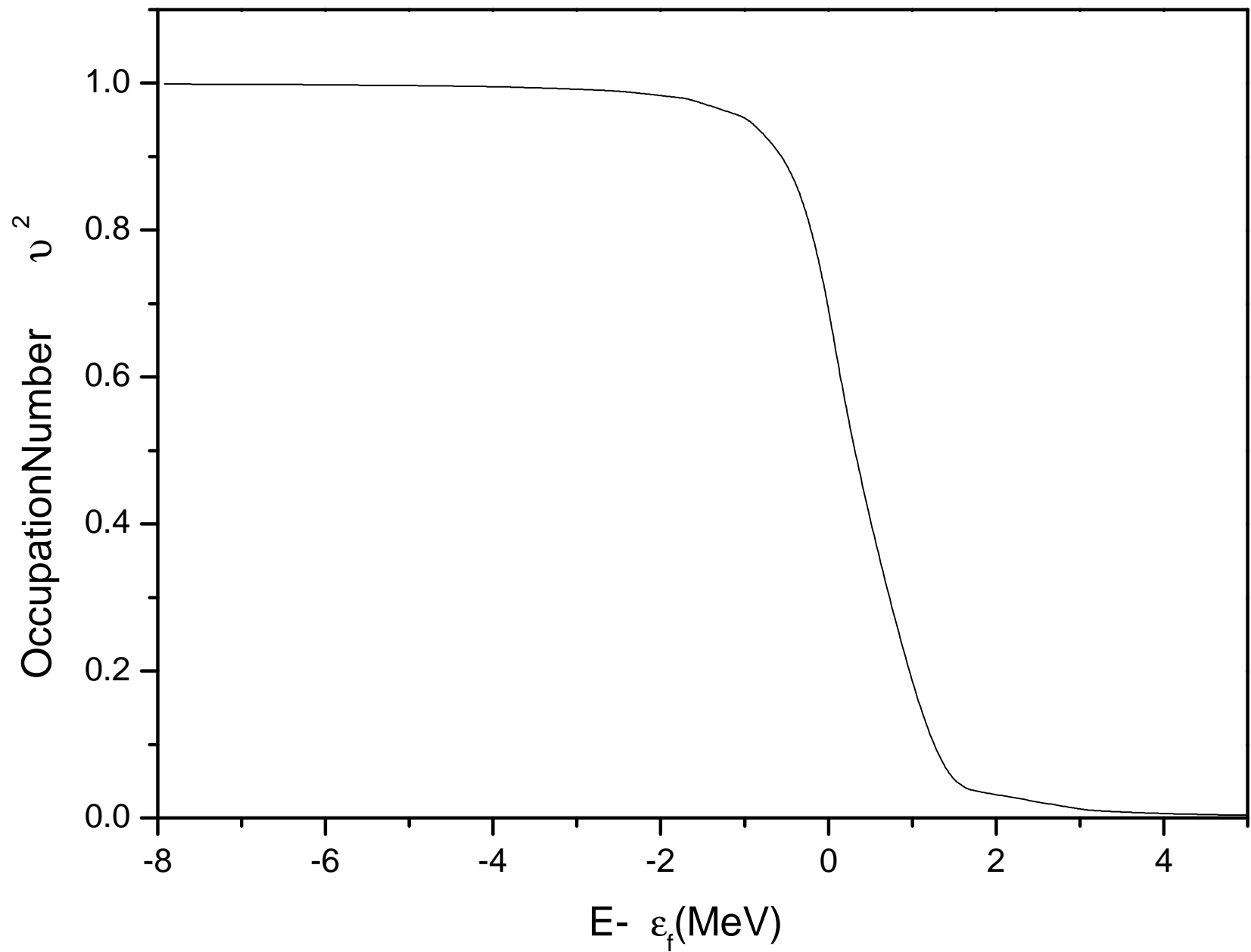
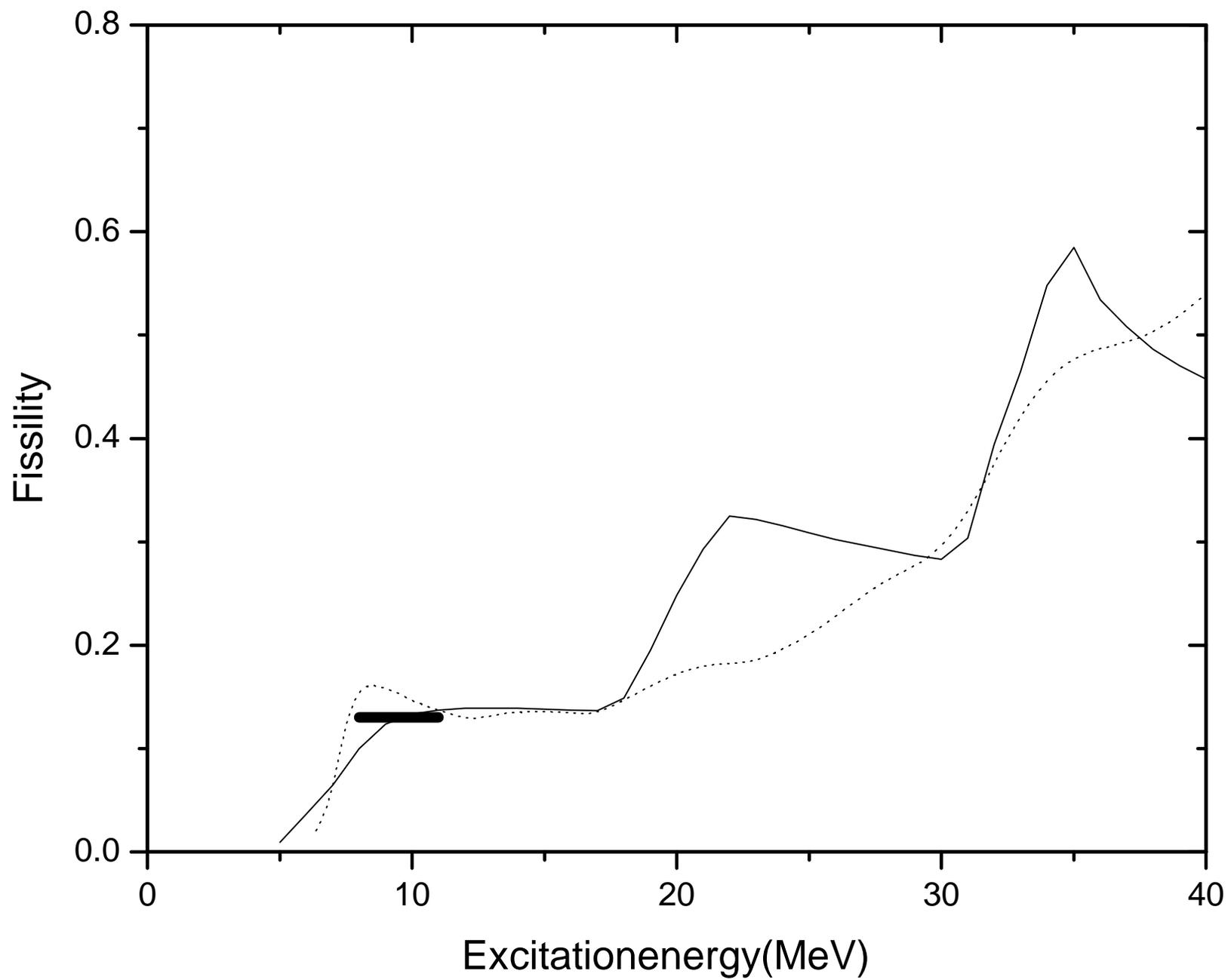
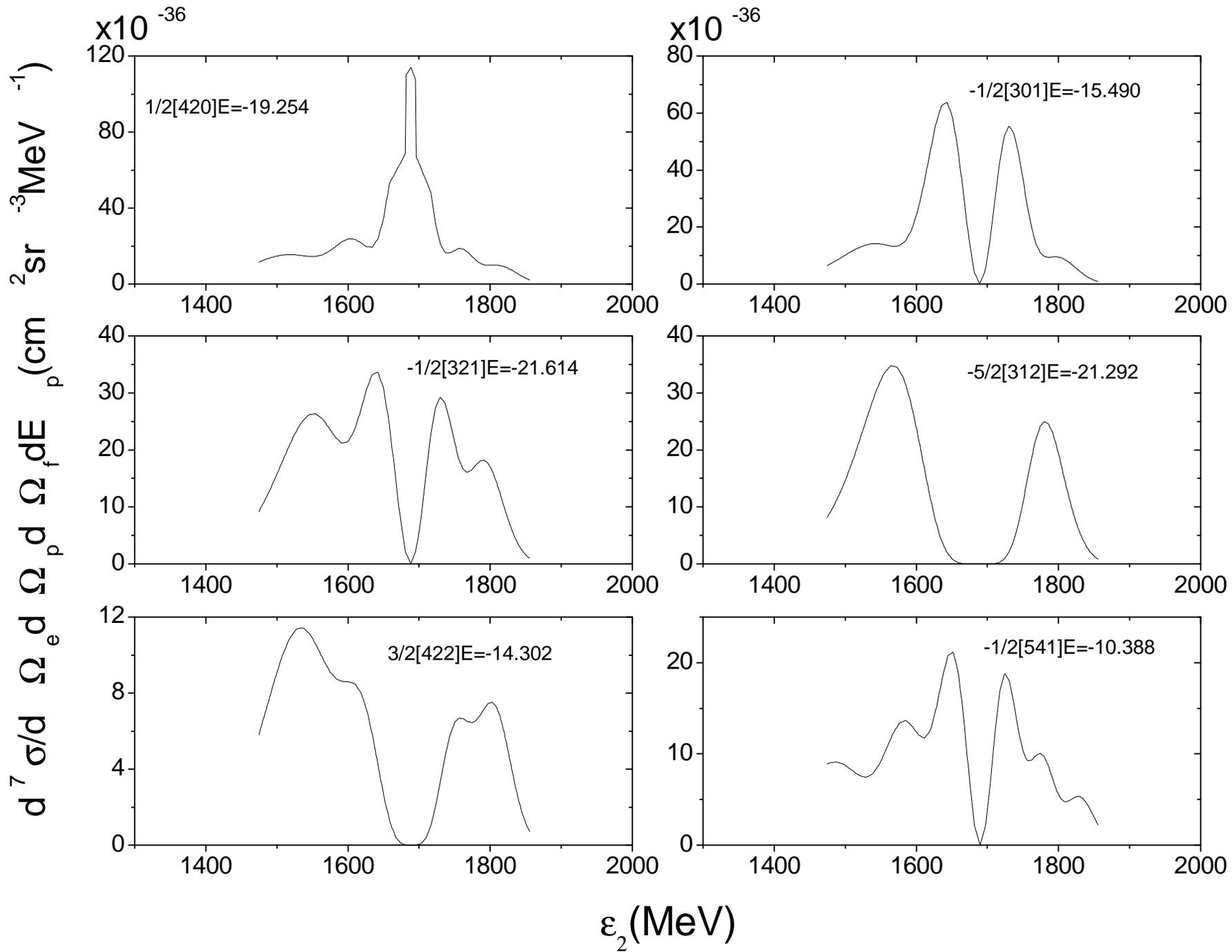
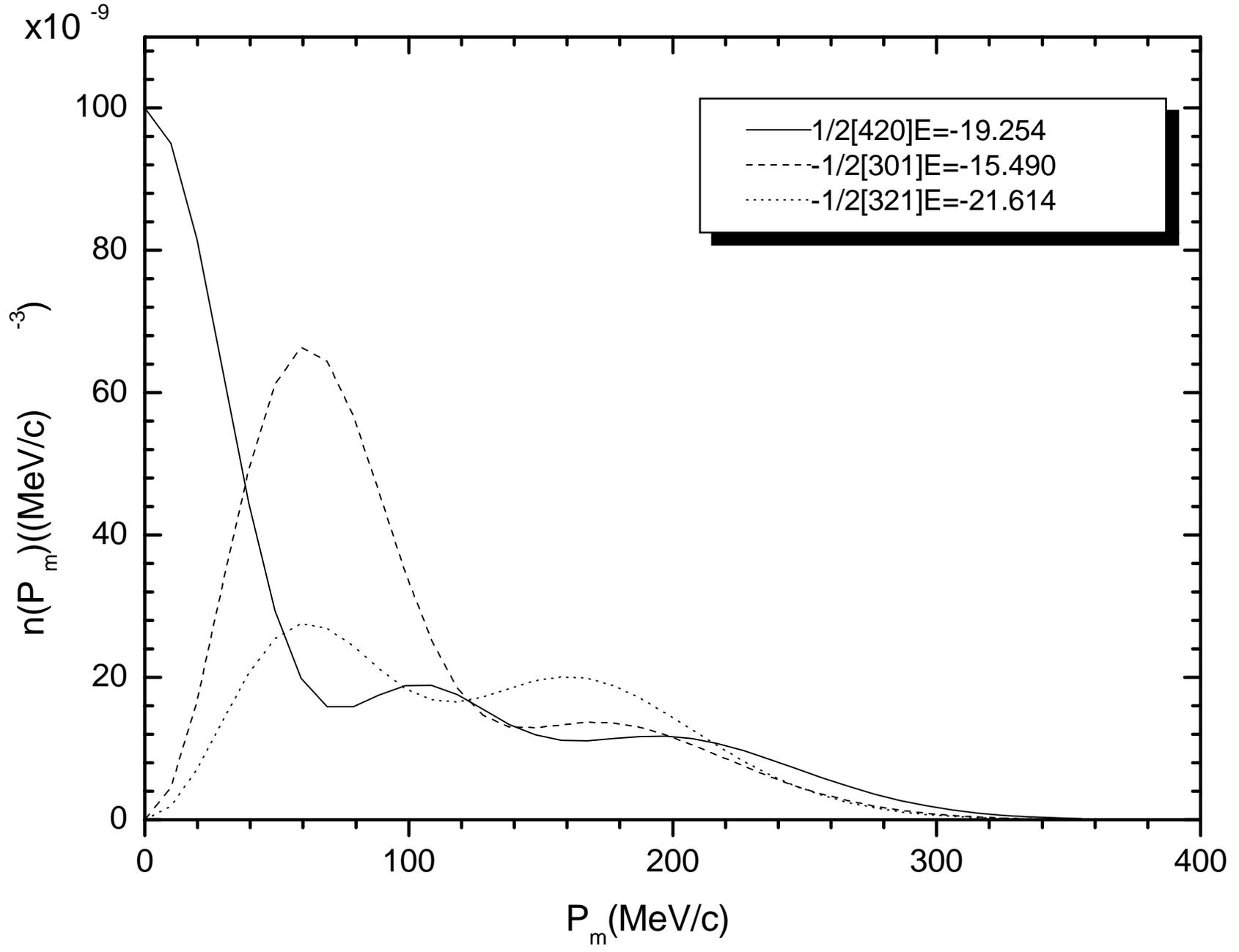
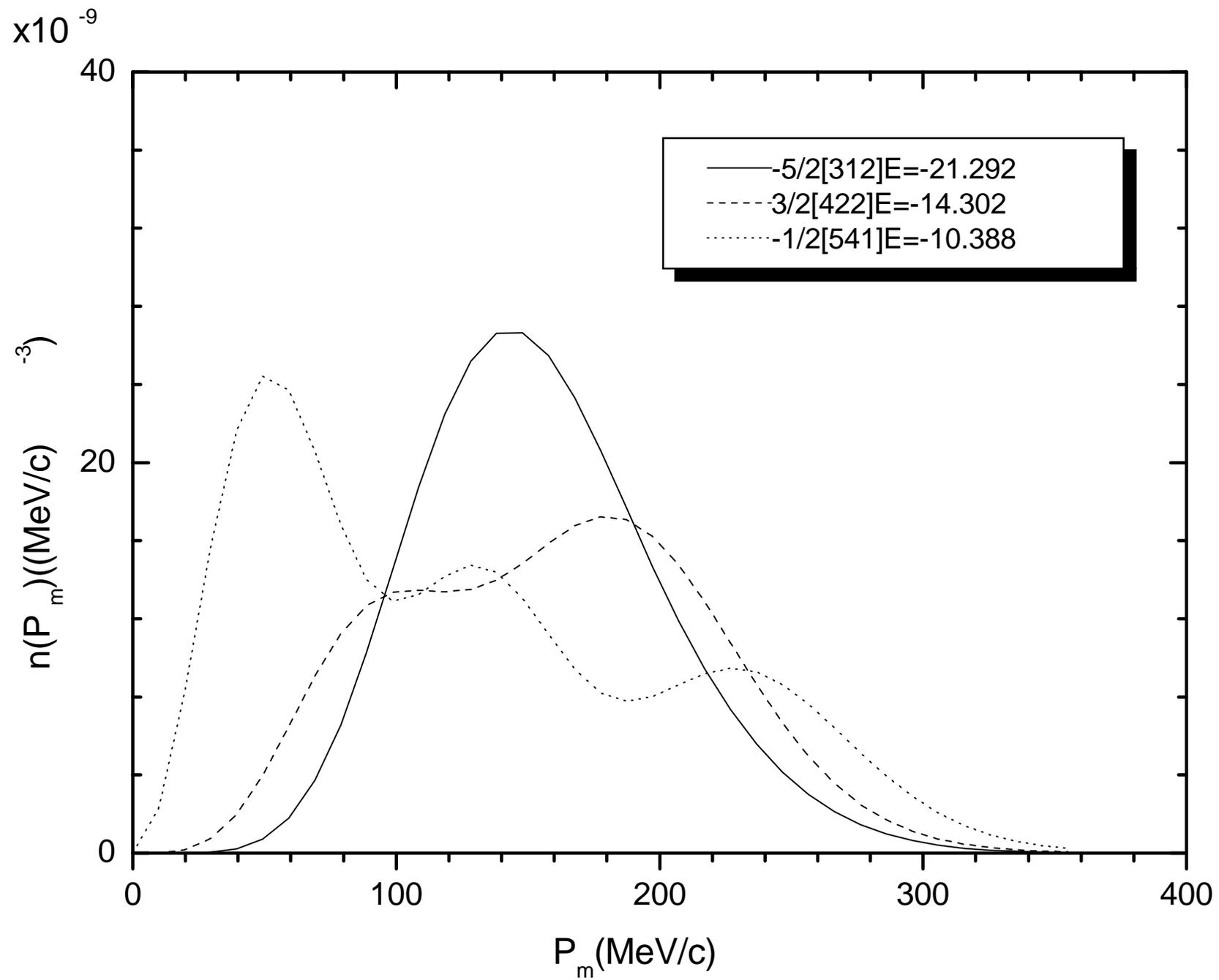
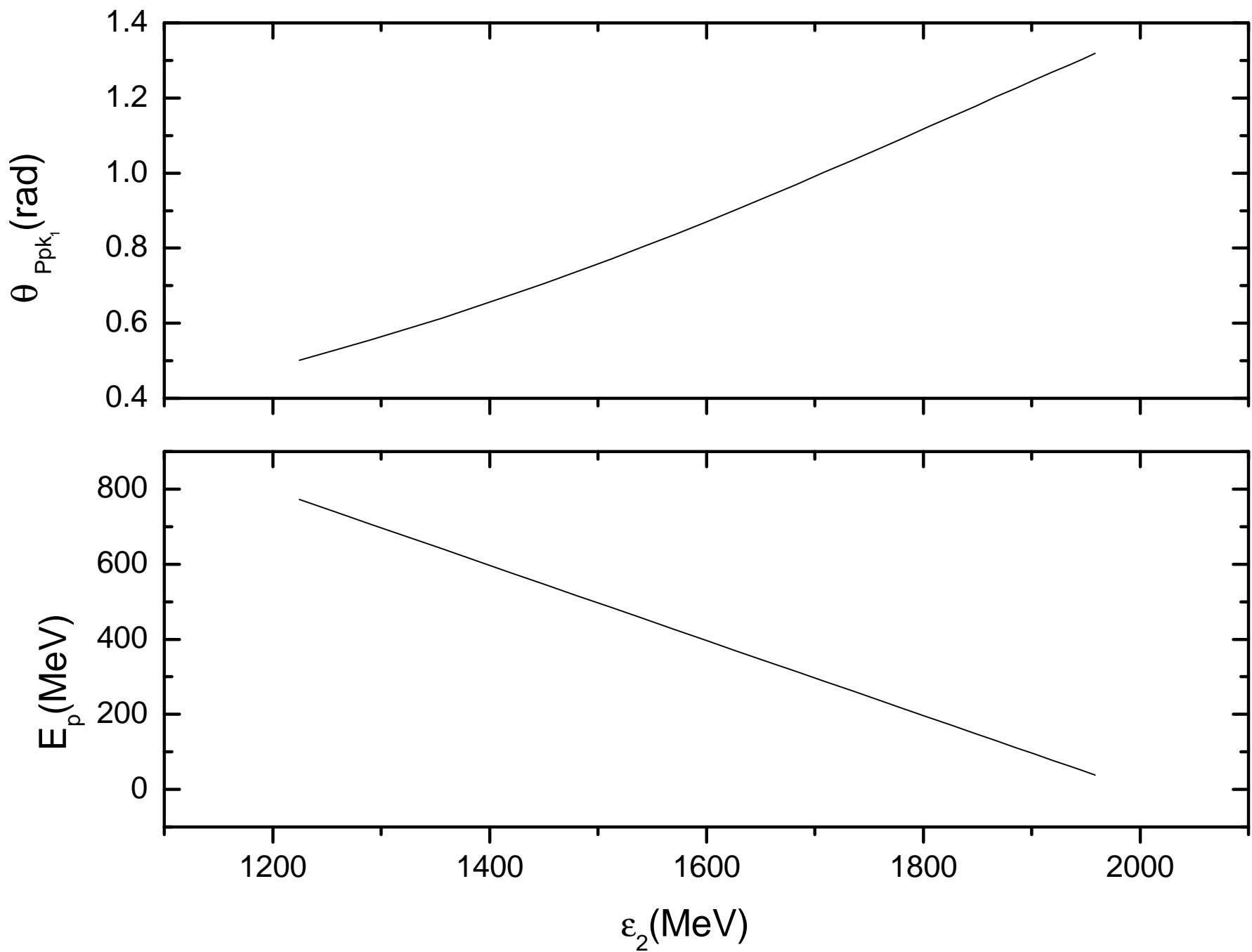
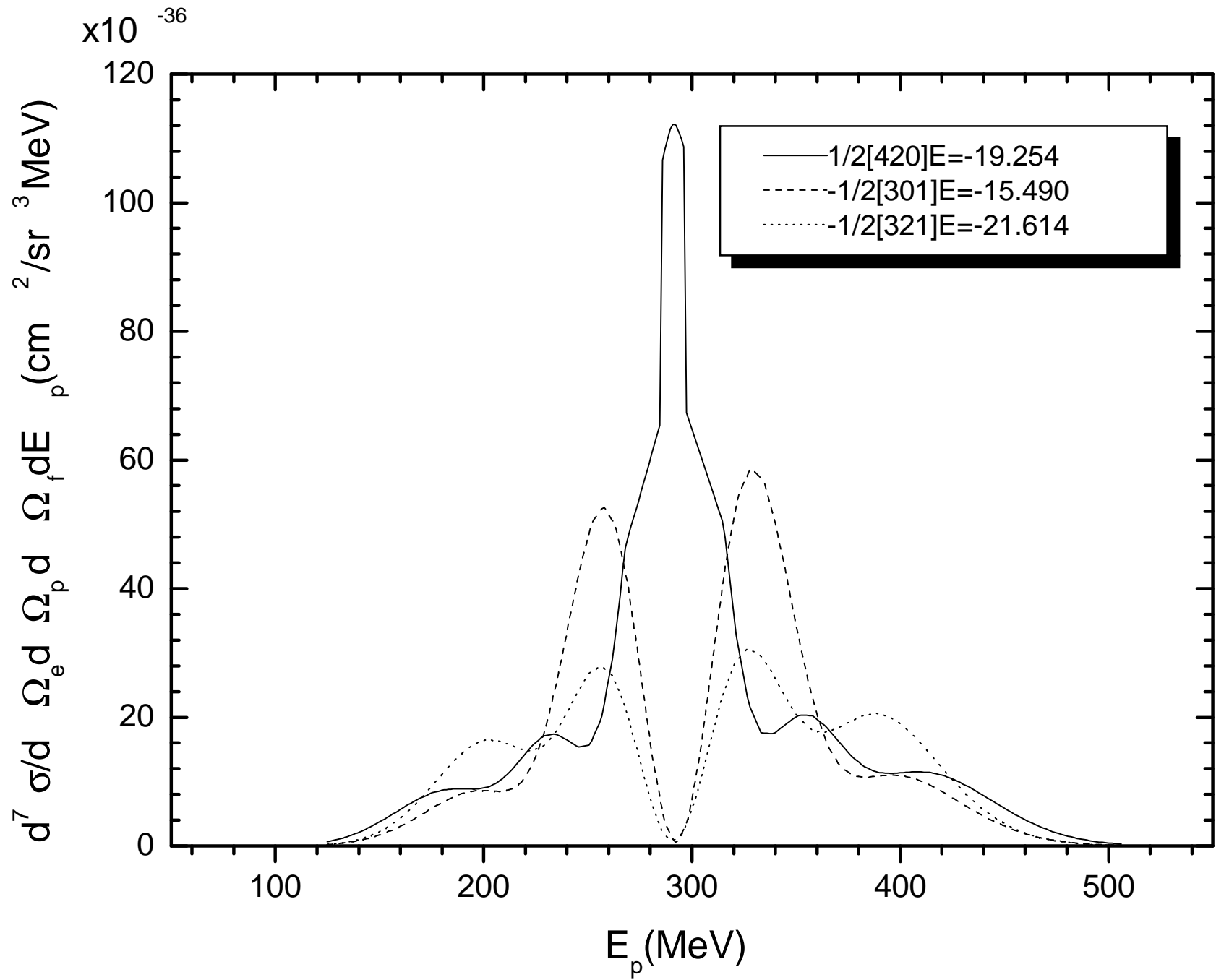
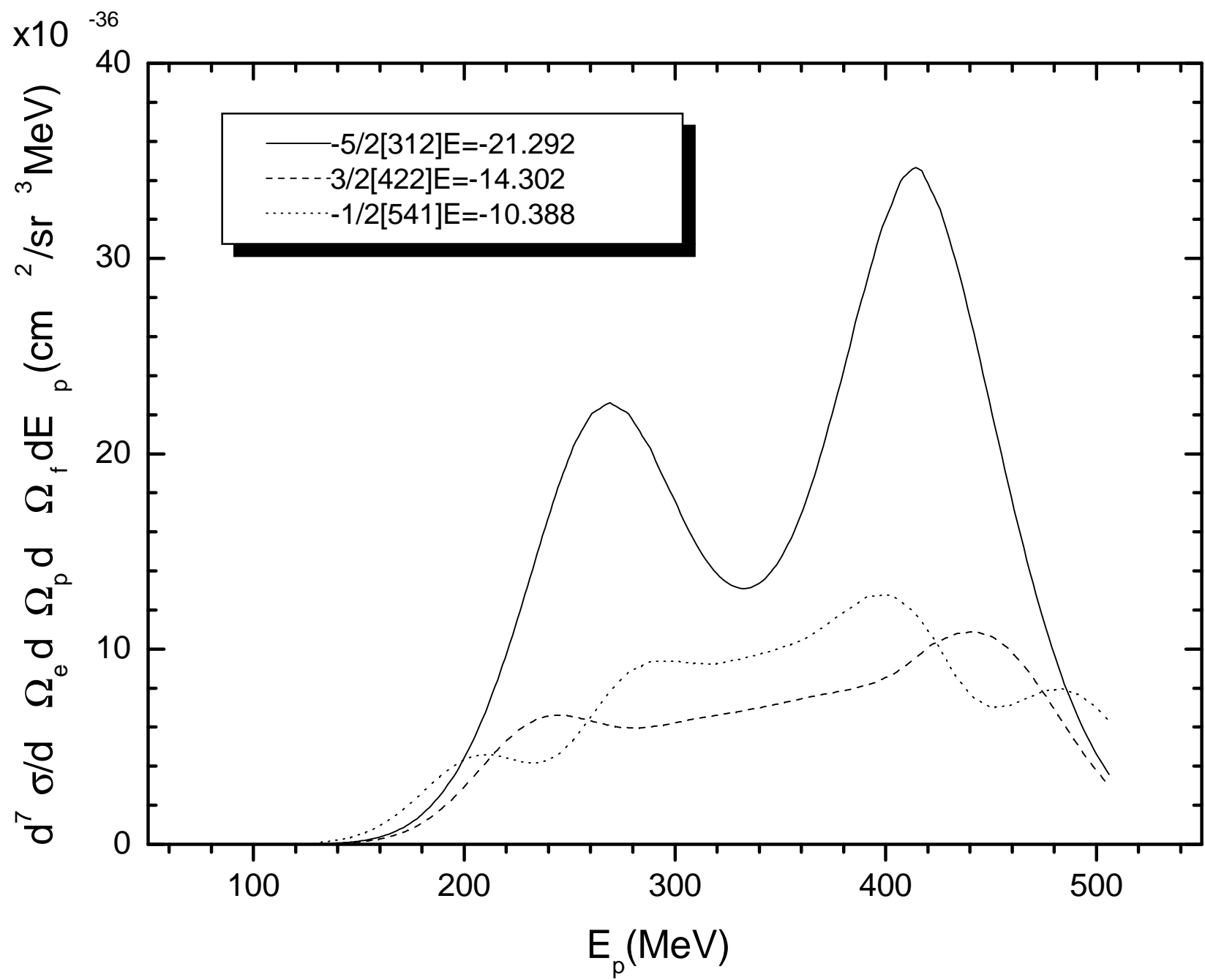
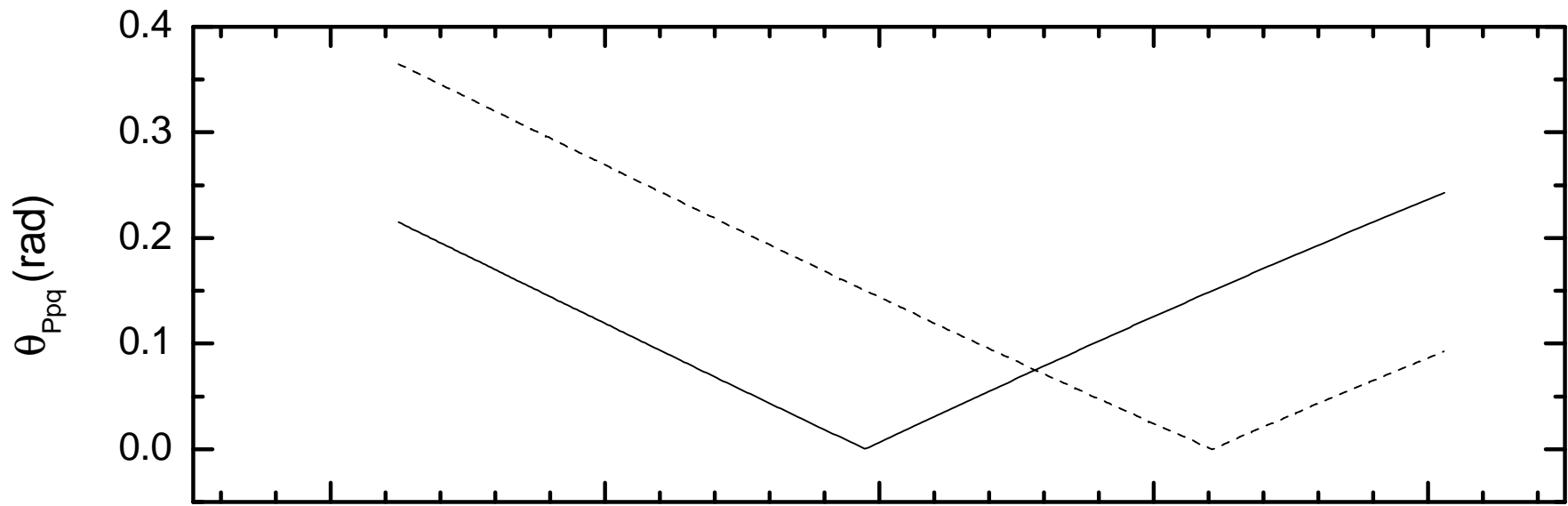
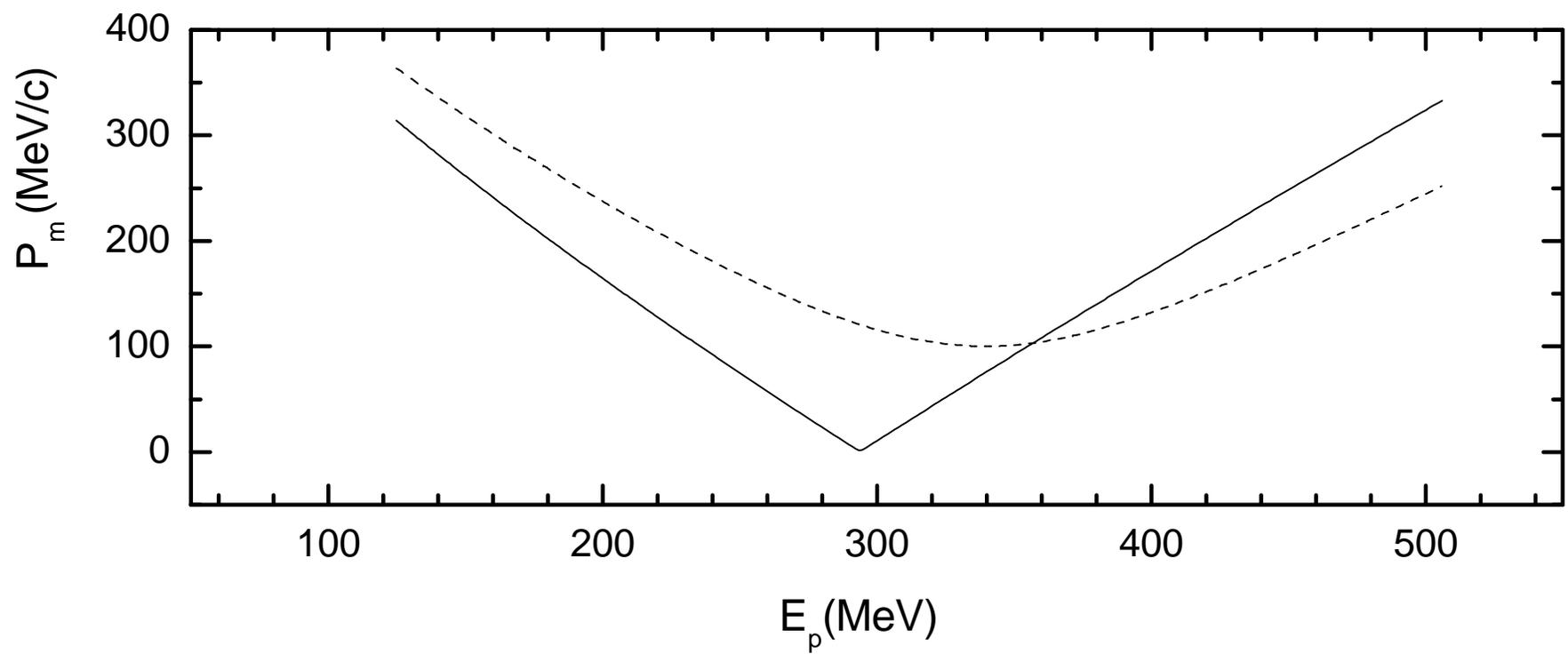
\end{document}